# Imaging across the Short-Wave Infra-Red (SWIR) Band via a Flat Multilevel Diffractive Lens

**Monjurul Meem,**[1, ¥] **Sourangsu Banerji,**[1, ¥] **Apratim Majumder,**[1] **Curt Dvonch,**[2] **Berardi Sensale-Rodriguez,**[1] **and Rajesh Menon**[1, 3*]

[1] *Department of Electrical and Computer Engineering, University of Utah, Salt Lake City, UT 84112, USA.*
[2] *Collins Aerospace, Princeton, NJ 08545, USA. Currently with Lockheed Martin Corp.*
¥ *denotes equal contribution*
*Corresponding author: rmenon@eng.utah.edu*

**We designed and fabricated a flat multi-level diffractive lens (MDL) that is achromatic in the SWIR band (875nm to 1675nm). The MDL had a focal length of 25mm, aperture diameter of 8.93mm and thickness of only 2.6μm. By pairing the MDL with a SWIR image sensor, we also characterized its imaging performance in terms of the point-spread functions, modulation-transfer functions, and still and video imaging.**

Imaging in the Short Wave Infrared (SWIR) band is useful in myriad applications including monitoring of manufacturing processes, automobile driver assistance, precision agriculture, plastics recycling and biomedical imaging **[1]**. The primary driver for the utilization of SWIR cameras has been advancements in the image sensors. However, the conventional refractive lenses used in these systems can often be cumbersome due to their thickness, weight and presence of chromatic aberrations. Combining Diffractive optical elements (DOEs) with refractive lenses have been proposed for correcting chromatic aberrations for many years **[2, 3]**, and demonstrated with the advent of semiconductor lithography **[4]**. It is mistakenly thought that diffractive lenses are unable to correct chromatic aberrations on their own. On the contrary, we have demonstrated achromatic multi-level diffractive lenses (MDLs) in the visible, **[5-7]** in the LWIR, **[8]** and in the THz **[9,10]** bands with excellent imaging performance. Surprisingly, we also showed that MDLs can potentially enable imaging over a practically unlimited operating bandwidth **[11]**.

In this Letter, we experimentally demonstrate a single MDL with a diameter of 8.93mm designed to be achromatic in the short-wave infrared regime, λ=875nm to 1675nm with focal length = 25mm and NA = 0.17 (Fig. 1a). The MDL is comprised of concentric rings of equal width (3μm), but varying heights (between 0 and 2.6μm, and up to 100 discrete levels). The distribution of heights is selected by maximizing the wavelength averaged focusing efficiency, which is computed as the ratio of power inside 3 times the full-width at half-maximum (FWHM) in the focal plane to the total incident power **[9]**.

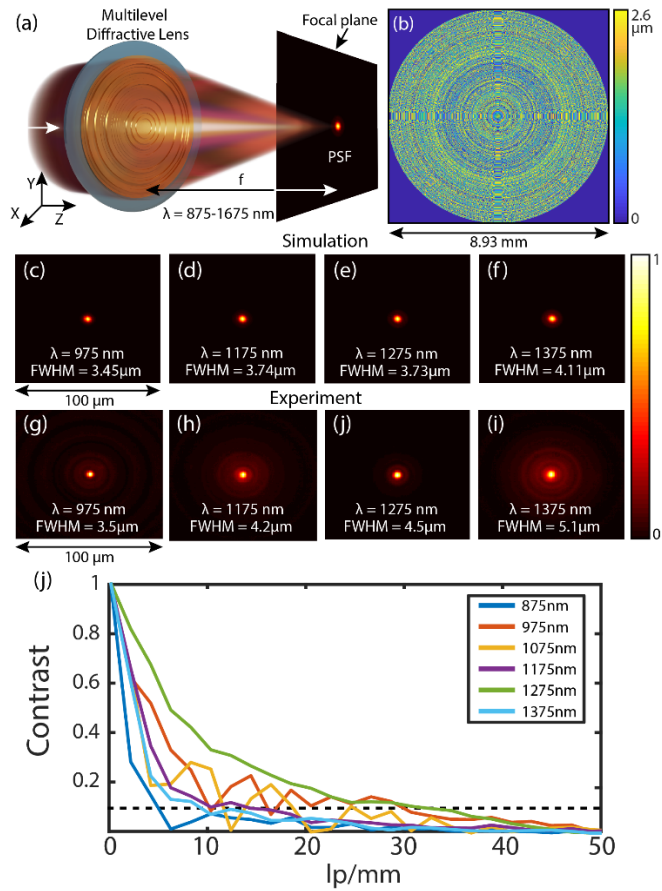

**Fig. 1. (a)** Schematic of the SWIR MDL. **(b)** The optimized height profile with focal length = 25mm and NA = 0.17. **(c-f)** simulated and **(g-i)** measured point-spread functions at the λ= 975 nm, 1175 nm, 1275 nm and 1375 nm, respectively. **(j)** Modulation transfer function (MTF) of the broadband MDL at λ = 875 nm – 1375 nm was obtained by taking a Fast Fourier Transform (fft) of the measured PSFs.

The dispersion of a positive-tone photoresist (S1813, Microchem) was also assumed [**5**]. The operating bandwidth was 875nm to 1675nm, which was dictated by the quantum efficiency of the InGaAs image sensor [**12**]. The 320CSX SWIR camera provided by Sensors Unlimited, Inc., a Collins Aerospace company, includes a 320X256 pixel InGaAs focal plane array with a 12.5μm pixel pitch, wide dynamic range and linear pixel response. As has been described previously [**8**], a gradient-assisted direct-binary-search algorithm was applied to perform the optimization of the MDL geometry. The optimized height distribution of the rings is shown in Fig. 1(b). The simulated point-spread functions (PSFs) are shown in Figs. 1(c)-(f) for λ=975nm, 1175nm, 1275nm and 1375nm, respectively. The PSFs were also measured by illuminating the MDL with the collimated beam from a super-continuum source (SuperK Extreme, NKT Photonics) equipped with a tunable filter. The image sensor was placed in the focal plane, while the illumination wavelength was changed. The captured images of the PSFs after dark frame subtraction are shown in Figs. 1(g)-(i). Note that our super-continuum source cuts of for wavelengths above ~1375nm, so we were not able to characterize the PSFs for wavelengths longer than that.

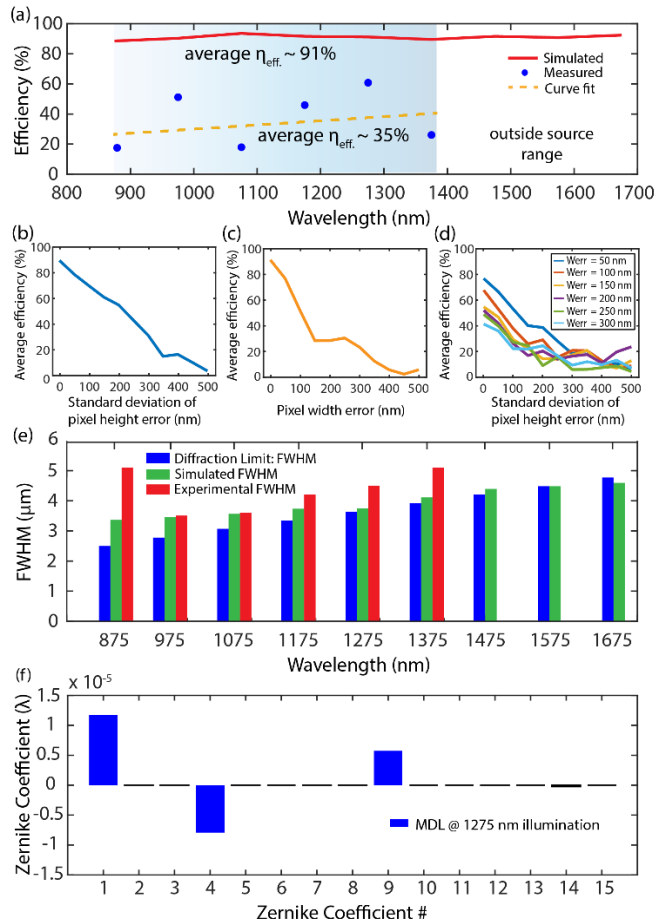

**Fig. 2. (a)** Simulated and measured focusing efficiency as a function of wavelength. A systematic study of the fabrication errors with relation to **(b)** standard deviation of the pixel height error, **(c)** pixel width error and **(d)** effect of both pixel width and standard deviation of the pixel height error. The effect of error in pixel width is more prominent among the two with the combined effect bringing down the error tolerance to ~100 nm for ~50% efficiency. **(e)** Full-width at half-maximum (FWHM) as a function of wavelength. **(f)** Simulated Zernike coefficients of the MDL at λ = 1275nm.

The focusing efficiency was simulated as described earlier [**7**] and computed from the measured PSFs. The results are summarized in Fig. 2(a). The simulated and measured (based on curve fitting) average efficiency for λ=875nm to 1375nm were 91% and 35%, respectively. To investigate this large discrepancy between simulation and experiment, we simulated the impact of ring-width errors (Fig. 2b), ring-height errors (Fig. 2c) and a combination of the two (Fig. 2d). The analyses suggest that with a ring-height error of ~100nm and ring-width error of ~400nm, the average focusing efficiency drops to ~35%. These errors are expected in our current fabrication process. However, commercial semiconductor manufacturing can attain errors far smaller than these can and hence, high efficiencies are possible. The full-width at half-maximum (FWHM) of the focal spots were computed for each design wavelength (Fig. 2e). Finally, utilizing the simulated wavefront from the MDL, we computed the equivalent lens aberrations in terms of Zernike coefficients (Fig. 2f), confirming relatively low aberrations.

As has been reported previously [**5-7**], the MDL was fabricated in a positive-tone photoresist (S1813) atop a glass substrate via grayscale lithography (MicroPG101, Heidelberg Instruments). Photographs and optical micrographs of the fabricated MDL are shown in Figs. 3(a)-(c). In order to characterize the imaging performance of the MDL, we imaged the Air Force resolution chart (Fig. 3d) and the Macbeth color chart (Fig. 3e). The illumination was via a halogen lamp. The distance between the MDL and the image sensor was ~26mm, while the distance between the object and the MDL was ~500mm. Conventional wiener deconvolution was applied to improve the quality of the images. The lines of group 1-4 corresponding to a spatial frequency of 2.83 line-pairs/mm (spatial period of 446μm) were resolved.

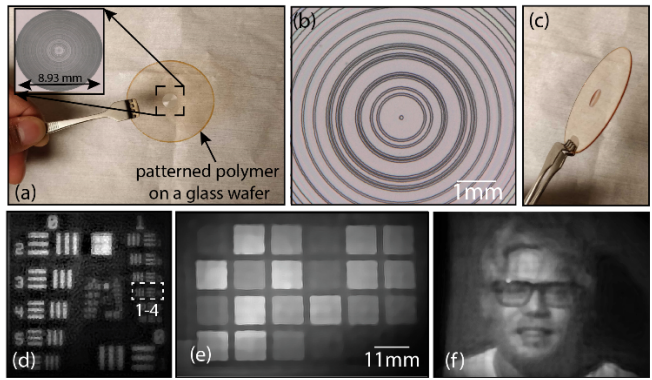

**Fig. 3. (a)** Photograph of the MDL (inset = optical micrograph). **(b)** Magnified optical micrograph of the central portion of the MDL. **(c)** Side view of the MDL. Images captured under ambient indoor lighting of **(d)** the Air Force resolution chart, **(e)** the Macbeth color chart and **(f)** a human face (Supplementary videos included). Wiener deconvolution was applied to these images.

Table 1. Summary of the reported work in comparison to previously reported thin flat metalenses.

| Material | Wavelength | Bandwidth | N.A. | Focal Length/ Diameter | Measured Efficiency | Feature Width/Height | Polarization | Reference |
|---|---|---|---|---|---|---|---|---|
| a-Si | 1.2μm-1.65μm | 450 nm | 0.13 | 800μm/200μm | ~32% | 100 nm/800nm | Polarization Insensitive | [13] |
| a-Si | 1.3μm-1.8μm | 500 nm | 0.04 | 7.5mm/600μm | 24%, 22%, 28% | 75 nm/600nm | Linear | [14] |
| Au/SiO2/Au | 1.2μm-1.68μm | 480 nm | 0.27 | 100μm/55.55μm | 12.44% | 40nm/30nm | Circular | [15] |
| Au/SiO2/Au | 1.2μm-1.68μm | 481 nm | 0.22 | - | 8.4% | 40nm/30nm | Circular | [15] |
| Au/SiO2/Au | 1.2μm-1.68μm | 482 nm | 0.32 | - | 8.56% | 40nm/30nm | Circular | [15] |
| Au | 0.532μm-1.08μm | 548 nm | - | 7μm/10μm | ~ 20% | 100nm,60nm, 40nm/40nm | Linear | [16] |
| polymer | 0.875μm – 1.675μm | 800 nm | 0.17 | 25mm/8.93mm | ~35% | 3μm/2.6μm | Polarization Insensitive | **This work*** |

Recently, metalenses have been touted as promising candidates for flat, lightweight lenses **[13-16]**. However, we showed that MDLs are far easier to fabricate and offer better imaging performance than metalenses **[17]**. In order to clarify this point further, we summarize the key characteristics of exemplary metalenses in the IR and compare these to our MDL in Table 1. There are two important points to note. Firstly, the bandwidth of our MDL (800nm) is far higher than any metalens demonstrated in the IR so far. Secondly, our MDLs require a minimum feature width of 3μm, which is far easier to achieve compared to the ~100nm feature widths required by metalenses. Furthermore, the MDL can be replicated inexpensively in polymers using imprint lithography, while metalenses typically require high-refractive-index materials.

In summary, we demonstrated good imaging performance using a single flat MDL in the wavelength range from 875nm to 1675nm (SWIR band) with diameter = 8.93mm, focal length = 25mm and NA=0.17. The thickness of the active region of our MDL is only 2.6μm and consequently its weight can be negligible as well. Such flat MDLs can replace multiple lens systems and enable far simpler, lighter and less expensive SWIR cameras.

**Funding.** Office of Naval Research grant N66001-10-1-4065.

**Acknowledgments**. We thank Brian Baker, Steve Pritchett and Christian Bach for fabrication advice, and Tom Tiwald (Woollam) for measuring dispersion of materials. We would also like to acknowledge the support from Amazon AWS (051241749381) for help with the computing facilities. We gratefully acknowledge Sensors Unlimited (Collins Aerospace) for lending us the InGaAs focal plane array.

**Disclosures**. Rajesh Menon is co-founder of Oblate Optics, Inc., which is commercializing technology discussed in this manuscript. The University of Utah has filed for patent protection for technology discussed in this manuscript.